# Física, investigación científica y sociedad en la Argentina de 1920-1930

**Alejandro Gangui** | Instituto de Astronomía y Física del Espacio, CONICET - Universidad de Buenos Aires
gangui@iafe.uba.ar
https://orcid.org/0000-0002-4864-5348

**Eduardo L. Ortiz** | Imperial College, London
e.ortiz@imperial.ac.uk
https://orcid.org/0000-0001-8236-8770

**RESUMEN**
Analizamos las investigaciones científicas llevadas a cabo en el Instituto de Física de la Universidad Nacional de La Plata en la primera mitad del siglo XX, y el contexto cultural y social en el que estuvieron inmersas. Nos concentramos especialmente en las actividades desarrolladas por el físico argentino Ramón G. Loyarte, quien fue una personalidad emblemática del mundo científico, educativo, cultural y político de la Argentina en esos años. Discutimos sus trabajos más trascendentes en física experimental y mecánica cuántica, sus actividades de gestión y promoción de la ciencia y el impacto internacional de sus propuestas científicas, así como también el origen de las polémicas que desataron sus ideas más osadas. Para estos últimos tópicos empleamos una herramienta novedosa: examinamos los comentarios sobre sus trabajos publicados en prestigiosas revistas internacionales de reseña científica, que ayudan a comprender de manera más cabal y contemporánea los descubrimientos de Loyarte.

**Palabras clave** Instituto de Física de La Plata - Ramón G. Loyarte - ciencia y política - hipótesis de un nuevo cuanto de energía rotacional - revistas internacionales de reseña científica.

**ABSTRACT**
*We analyse the scientific research carried out at the Institute of Physics of the National University of La Plata in the first half of the 20th century, and the cultural and social context in which they were immersed. We focus especially on the activities carried out by the Argentine physicist Ramón G. Loyarte, who was an emblematic personality in the scientific, educational, cultural and political world of Argentina in those years. We discuss his most important works in experimental physics and quantum mechanics, his activities in the management and promotion of science and the international impact of his scientific proposals, as well as the origin of the controversies unleashed by his most daring ideas. For the latter topics we employ a novel tool: we examine the comments on his work published in prestigious international scientific review journals, which help to understand Loyarte's findings in a more comprehensive and contemporary way.*

## Introducción

En este trabajo nos ocupamos de las investigaciones desarrolladas bajo la supervisión de Ramón G. Loyarte, físico argentino que fue director del Instituto de Física de la Universidad Nacional de La Plata (UNLP) y personalidad destacada en la ciencia de la Argentina en la primera mitad del siglo XX. Loyarte fue miembro de la Academia Nacional de Ciencias de Buenos Aires en el período inicial de sus actividades y fue el primer físico argentino galardonado con el Premio Nacional de Ciencias. Cuando aún estaba por cumplir los cuarenta años, fue elegido presidente de la Universidad de la Plata, a la que se había acercado años antes como alumno.

En sus tareas como investigador Loyarte inicialmente continuó con líneas de investigación introducidas en el Instituto de Física por el último de sus directores alemanes: Richard Gans. Esos estudios se relacionaban, principalmente, con aspectos de la moderna teoría del magnetismo, en la que se comenzaban a tener en cuenta los aportes teóricos de la nueva física cuántica. Poco después del regreso de Gans a Alemania, Loyarte publicó una serie de estudios en los que, en alguna forma, trataba de verificar experimentalmente ideas introducidas por la mecánica cuántica. Los resultados de algunas de sus investigaciones, o de las suyas junto con algunos de sus colaboradores locales, fueron publicados en revistas científicas extranjeras de amplia circulación.

En este trabajo nos concentramos en el lugar que tuvo el Instituto de Física de La Plata en el contexto de la ciencia argentina, y en los esfuerzos realizados por varios actores, Loyarte entre ellos, para la promoción de las ciencias en ese país en la década de 1920. A continuación, discutimos los mecanismos de comunicación nacional e internacional que fueron empleados por los físicos de la UNLP para difundir los resultados de sus investigaciones. Destacamos los trabajos de Loyarte y de sus colaboradores en la segunda mitad de la década de 1920, y cuál fue su impacto científico en el extranjero. Analizamos entonces la recepción que tuvo su propuesta sobre la rotación cuantificada de los átomos, es decir, sobre la posible existencia de un nuevo cuanto de energía rotacional, en medios académicos fuera de la Argentina mediante la consulta de revistas internacionales de referencia, que emplearemos como una herramienta de evaluación novedosa.

En una continuación de nuestro estudio (Gangui y Ortiz, 2022.1) veremos con cierto detalle los conflictos que surgieron en el seno del Instituto de Física de La Plata debido a la crítica que uno de los miembros jóvenes de ese Instituto hizo de los trabajos y propuestas de Loyarte. Esta *polémica* excedió los confines de la Argentina y se extendió por los medios académicos de Norteamérica y de Europa. También detallaremos la actividad de Loyarte en la política partidaria y en la política de la ciencia en la Argentina de los años 1930-1940. Como miembro del Parlamento Nacional, fue el encargado de promover interesantes iniciativas, algunas relacionadas con el avance de la ciencia en la Argentina. Por último, también veremos cómo Loyarte entró de lleno en el debate sobre la nacionalización de la dirección de institutos científicos en ese país.

## 1. El Instituto de Física de La Plata en el cuadro de la ciencia argentina

### 1.1 El Instituto de Física de La Plata: Ricaldoni, Bose, Gans

Como resultado de un movimiento de opinión iniciado hacia mediados de 1889 por el Senador Rafael Hernández (1840-1903), hermano del conocido poeta José Hernández (1834-1886), autor del *Martín Fierro*, en enero de 1890 se sancionó una ley que autorizaba la fundación de una universidad en la nueva ciudad de La Plata, creada como capital de la Provincia de Buenos Aires luego de la capitalización de la ciudad de Buenos Aires. Esa institución, de carácter provincial, tenía un sentido diferente del de las dos universidades nacionales que existían en ese momento en Argentina: la de Córdoba, fundada por los Jesuitas en 1613, en tiempos de la Colonia, y la de Buenos Aires, fundada dos siglos más tarde, en 1821, pocos años después de declarada la Independencia.

Hacia 1897 la nueva Universidad Provincial de La Plata fue organizada en base a cuatro disciplinas, representadas por cuatro facultades eminentemente profesionales: de Derecho, de Ciencias Médicas, de Química y Farmacia, y de Ciencias Fisicomatemáticas. A partir de entonces la institución platense vivió una existencia precaria, limitada por recursos reducidos que determinaron un funcionamiento intermitente. Consecuentemente encontró serias dificultades para atraer alumnos, que seguían prefiriendo la seguridad de las universidades tradicionales.

El 12 de agosto de 1905 el ministro de Instrucción Pública, Dr. Joaquín V. González (1863-1923), que era una figura nacional de relieve intelectual, sometió al Congreso Nacional un proyecto que concedía a aquella institución un carácter administrativo más amplio: el de *Universidad Nacional*. Al mismo tiempo, propuso una estructura inspirada en los antiguos *Colleges* ingleses, cuyo modelo había sido ya transportado, y enriquecido, a los Estados Unidos. Estas características la diferenciaban profundamente de las dos universidades anteriores. En esa nueva institución la ciencia, la enseñanza experimental y práctica, lo mismo que el régimen de internado, jugaban un rol central: su moto es expresivo, *Scientia et Patria* (Ortiz y Rubinstein, 2009; Ortiz, 2011).

El plan de González preveía la incorporación a la nueva Universidad Nacional de un grupo de valiosos institutos científicos, fundados en La Plata con anterioridad: un bien dotado Observatorio Astronómico, un rico Museo de Ciencias Naturales y una Escuela Práctica de Agronomía; ésta última establecida en el pueblo de Santa Catalina, en las cercanías de la ciudad de La Plata. El primero de esos institutos se anexó a la antigua Facultad de Ciencias Fisicomatemáticas, el segundo a la Facultad de Química y Farmacia y el tercero dio origen a la Facultad de Agronomía y Ciencias Veterinarias (Nazar Anchorena, 1927, pp. 104, 199).

Más tarde se agregó a la Universidad un Instituto de Física cuya organización y dirección se confió al ingeniero Tebaldo[1] J. Ricaldoni (1861-1923). A nivel de la enseñanza secundaria Ricaldoni era uno de los muy pocos profesores de física idóneos de quienes disponía el país en esos años; sin duda, el más activo, entusiasta y original. Además, contaba con una amplia experiencia práctica de laboratorio -y sobre todo de taller- que había demostrado enseñando física en el Colegio Nacional de Buenos Aires. En esos años la física era apreciada en la Argentina como una ciencia pura, pero también por sus fascinantes aplicaciones. Esta última percepción posiblemente no había estado ausente en la designación de Ricaldoni, conocido por sus inventos y por sus celebrados intentos en los campos de la telegrafía sin hilos y los rayos X.

A este primer director se le encargó la selección y adquisición de una colección de casi tres mil instrumentos de demostración. Ricaldoni los encargó a la fábrica de Max Kohl, AG, en Chemnitz, Alemania que, en esos años, era una de las firmas más conocidas y acreditadas en el campo de la manufactura de instrumentos científicos para la enseñanza. Los equipos de Kohl se utilizaban en escuelas de diferentes países de Europa y de los Estados Unidos. En La Plata, el material adquirido se utilizaría para ilustrar más elocuentemente las clases de física, y para darles un mayor sentido de modernidad. El costo de esa colección de instrumentos fue elevado.

En 1909 Ricaldoni fue reemplazado en la dirección del Instituto de Física por un físico contratado especialmente en Alemania, el Dr. Emil Bose (1874-1911), un joven discípulo del celebrado profesor Walther Nernst (1864-1941). Bose se incorporó al país con Margrete Elisabet Heiberg de Bose (1865-1952) que era su esposa y también su asistente (Pyenson, 1985; Bibiloni, 2000; Hunter y Pyenson, 2005). Con la designación de Bose las autoridades universitarias esperaban transitar de la mera transmisión de las ideas de las ciencias físicas a la creación local de nuevos conocimientos dentro de esa disciplina. Durante el periodo en el que Bose ejerció la dirección del Instituto, además de la incorporación de nuevos investigadores extranjeros, se dieron pasos fundamentales para asegurar el entrenamiento de estudiantes universitarios en ramas modernas de la ingeniería; en particular, para la creación de las carreras de Ingeniería Eléctrica y Mecánica (Pyenson, 1985; Ortiz, 2021).

---

[1] Teobaldo en algunos documentos.

Una de las primeras tareas de Bose fue trasladar su Instituto, que funcionaba en dependencias modestas, a un edificio que inicialmente había sido designado como Gabinete de Física y Química del Colegio Nacional asociado a la Universidad de La Plata, que Bose remodeló completamente. La UNLP aceptó una proposición de Bose de adquirir, nuevamente, una segunda e importante colección de instrumentos de laboratorio (Ortiz, 2021). Este nuevo grupo de instrumentos estaba orientado a posibilitar investigaciones científicas originales, en las que se preveía que comenzarían a participar algunos estudiantes locales. Al mismo tiempo, las dos series de instrumentos se utilizaron para ilustrar con experimentos las clases universitarias de física; sin duda, a un nivel más avanzado que en el pasado. Ese instrumental se utilizó también para ilustrar conferencias y demonstraciones dirigidas a un público general educado. Esas veladas tuvieron un impacto muy considerable sobre la comunidad local y ayudaron a consolidar la posición del Instituto, y también de las ciencias físicas, dentro del ambiente no-científico local. Los comentarios favorables insertados en diarios de circulación local y nacional ayudaron a dar a esas veladas, y también a la física, una nueva perspectiva.

Debido a deficiencias de las empresas técnicas locales, Bose y sus alumnos se vieron forzados a participar, muy directamente, en algunos aspectos técnicos de la instalación de los nuevos laboratorios. Por ejemplo, tomaron a su cargo el montaje de algunos equipos complejos, lo mismo que el tendido de las instalaciones de agua, provisión de energía eléctrica, gas, vacío, etc. (Heiberg de Bose, 1911). Esas tareas, totalmente imprevistas, contribuyeron a desarrollar la habilidad manual y la pericia técnica de los jóvenes colaboradores locales.

Lamentablemente, Bose contrajo tifus en La Plata y falleció en mayo de 1911. Poco más tarde la UNLP, y el país en general, comenzó a transitar por un período de graves restricciones financieras. La reducción del presupuesto universitario llegó a hacer peligrar el pago del salario de sus profesores. Sin embargo, una vez superada esta etapa y luego de diversas alternativas, la UNLP logró contratar a otro destacado físico alemán: Richard Gans (1880-1954), en reemplazo de Bose. Gans era un físico formado en algunas de las mejores escuelas científicas de Alemania, con una extensa experiencia y renombre como investigador científico original. En La Plata, donde llegó en 1912, contribuyó a ampliar, considerablemente, el primer grupo de estudiantes de física nucleado por Bose.

En 1914 Nernst visitó la UNLP y, luego de su viaje, tres egresados: José B. Collo (1887-1968), Teófilo Isnardi (1890-1966) y Ramón G. Loyarte (1888-1944) recibieron becas externas financiadas por el Estado argentino que les permitieron trasladarse a Alemania. Allí estudiaron bajo la supervisión de Nernst, Max Planck (1858-1947) y otros distinguidos científicos alemanes.

Durante la dirección de Gans esos tres físicos argentinos recibieron su doctorado en la UNLP. En 1925 otro estudiante, Enrique Loedel Palumbo (1901-1962), de origen uruguayo y más joven que los anteriores, obtuvo también su doctorado en física (Gangui y Ortiz, 2020).

En la primera mitad del siglo XX ese pequeño grupo de científicos jugó un papel singular en el desarrollo de su disciplina en la Argentina.

En 1925, a pesar de las muy difíciles condiciones económicas en las que se encontraba Alemania, Gans decidió dejar la dirección del Instituto y aceptar una cátedra de física en la Universidad de Königsberg (Gaviola, 1950). Luego de algunas alternativas, su exalumno Loyarte fue elegido para sucederlo en la dirección del Instituto de Física. En La Plata los temas de investigación de los antiguos alumnos de Gans continuaron, por muchos años, dentro de la temática que éste había definido y que, entonces, no estaba alejada de las fronteras mismas de la investigación internacional en física.

## 1.2 La personalidad científica de Loyarte dentro del cuadro contemporáneo de la ciencia argentina

En este trabajo nos ocuparemos, más particularmente, de las actividades científicas de Ramón Godofredo Loyarte que, como ya hemos indicado, alcanzó una posición destacada en el mundo de la ciencia argentina, de la vida universitaria y, también, de la política nacional de su tiempo.

A comienzos del siglo Loyarte fue alumno del Colegio del Salvador, en Buenos Aires, (Furlong 1944, p. 684). En esos años este colegio tradicional, patrocinado por la Compañía de Jesús, hacía esfuerzos serios por modernizar su enseñanza de las ciencias (Ortiz, 2019). Más tarde, como ya hemos indicado, Loyarte fue alumno de la nueva UNLP, luego de la universidad de Göttingen y, a su regreso, obtuvo su doctorado en La Plata.

En 1916 Loyarte fue designado Profesor de Física Especial (una de las asignaturas de la Licenciatura en Física) en la UNLP (Archivo General, UNLP; Anónimo, 1944; Gangui y Ortiz, 2020). En 1924 el ministro de Educación, Dr. Antonio Sagarna (1874-1949), lo designó asesor científico de su departamento de estado y dos años más tarde lo incorporó a una Comisión Ministerial que se proponía contribuir a actualizar los contenidos científicos de los programas de la enseñanza secundaria. Las experiencias que Loyarte recogió en esas comisiones lo motivaron para redactar, él mismo, un nuevo texto de física elemental. Su obra estaba destinada a reemplazar, con un enfoque más moderno, tanto a los antiguos tratados de autores franceses, como los más recientes de Ricaldoni y de otros autores locales. Más importante aún, con esa nueva obra Loyarte se proponía contribuir a desplazar de la enseñanza una visión preferentemente descriptiva de la física, que hasta entonces había dominado en su enseñanza y que había comenzado a ser seriamente cuestionada por Bose (Ortiz, 2021). Loyarte invitó al joven físico Enrique Loedel Palumbo que, como él mismo, se había doctorado en física bajo la supervisión de Gans, a contribuir en la escritura de esa obra.

En paralelo con sus estudios científicos, quizás como una garantía de supervivencia para un graduado en una carrera nueva, Loedel Palumbo había cursado el Profesorado en Ciencias en la sección de Pedagogía de la UNLP; lo mismo habían hecho Loyarte y otros graduados en física. Sin duda, esta segunda formación académica, ganada en una Facultad definitivamente orientada hacia la pedagogía moderna, facilitó el éxito del texto de Loyarte y su colega (Loyarte y Loedel Palumbo, 1928; 1932). Esta obra de física elemental para estudiantes de la escuela secundaria alcanzó un gran número de ediciones (Gangui y Ortiz, 2022.2).

A principios de la década de 1920 la *Sociedad Científica Argentina* (SCA), que había sido fundada en 1872, decidió celebrar sus primeros cincuenta años de vida con la edición de una serie de monografías que analizaran el desarrollo de las diferentes ciencias en la Argentina entre 1872 y 1922. Diversos especialistas locales se ocuparon de la redacción de esas monografías: Loyarte fue invitado a encargarse de la que se ocupaba de las ciencias físicas. Resultado de sus esfuerzos fue la monografía *La evolución de la Física*, que apareció en Buenos Aires en 1924 (Loyarte, 1924).

En este interesante fascículo, de algo más de 80 páginas, Loyarte se ocupó de la evolución de esa ciencia en la Argentina en un período más amplio que el que requería la SCA. Comenzó considerando el estado de los estudios de física en la época colonial y luego se ocupó del período comprendido entre la independencia y la fundación de la Universidad de Buenos Aires, 1810 a 1821. Más tarde discutió brevemente las novedades introducidas en la enseñanza de la física por el eminente físico italiano Ottaviano Fabrizio Mossotti (1791-1863) que enseñó en la nueva Universidad de Buenos Aires entre 1827 y 1835. A continuación, se ocupó de los cambios ocurridos en la segunda mitad del siglo XIX y, finalmente, de la física en la Argentina en el período que va desde la fundación de la UNLP hasta principios de la década de 1920.

Para esa tarea, que realizó escrupulosamente, consultó las principales fuentes bibliográficas e históricas sobre la educación, la cultura y la universidad en la Argentina. En el capítulo sobre las dos primeras décadas del siglo XX hizo referencia a los cursos sobre física moderna dictados en la Facultad de Ciencias de Buenos Aires a partir de 1916 por el físico francés Camilo Meyer (1854-1918), radicado en la Argentina. Luego se ocupó de los cursos especiales 'sobre mecánica estadística, teoría de los "quanta" y de los fundamentos de la teoría de Bohr' que Gans dictó a sus colegas de La Plata en 1920 (Loyarte, 1924, pp. 77-78). Destacó, también, muy especialmente, el impacto que tuvo la visita del físico español Blas Cabrera (1878-1945), en 1920, sobre el avance de la física en la Argentina. Según Loyarte sus seminarios contribuyeron a actualizar temas contemporáneos de la física experimental y teórica.

Esta última visita fue facilitada por un convenio cultural celebrado entre la Universidad de Buenos Aires y la Institución Cultural Española de Buenos Aires. La Institución Cultural cubría los costos de ese programa, mientras que la Junta para Ampliación de Estudios colaboraba, desde Madrid, con una prudente selección de los profesores visitantes. Ese convenio posibilitó la visita de un grupo destacado de profesores españoles entre los que se puede citar, además de Cabrera, al matemático Julio Rey Pastor (1888-1962) y al filósofo José Ortega y Gasset (1883-1955): su influencia intelectual en la Argentina fue muy considerable (Ortiz, 1988).

Finalmente, Loyarte se refirió, brevemente, a los esfuerzos realizados por algunos de sus colegas contemporáneos. Con Ricaldoni no fue generoso; con su colega Teófilo Isnardi, autor de una obra reciente donde aconsejaba la inclusión de experimentos de física en la enseñanza de esa ciencia a nivel básico (Isnardi, 1913) expresó desacuerdos serios (Loyarte, 1924, pp. 79-80).

## 1.3 Loyarte, director argentino del Instituto de Física

Como ya hemos indicado, en 1925, luego de una intensa actividad científica en La Plata, Gans regresó a Alemania. Poco más tarde su exalumno Loyarte fue elegido para sucederlo en la dirección del Instituto de Física. Como veremos más adelante, además de su mérito científico, la elección de Loyarte para esa posición puede también percibirse como parte de una incipiente tendencia en favor de la nacionalización de la dirección de los grandes institutos de investigación de la Argentina.

Sin duda, algunos esperaban que los científicos contratados en el exterior, luego de su gestión en la dirección de institutos científicos nacionales, contribuirían también a la formación de discípulos locales capaces de continuar con su obra. Si bien éste había sido el caso en algunas instituciones científicas y técnicas locales, se podían citar numerosos ejemplos en los que ese objetivo, por una razón u otra, no se pudo cumplir. Un ejemplo citado a menudo en esos años fue el del Observatorio Nacional de Córdoba: después de su fundación quedó en manos de astrónomos estadounidenses por más de medio siglo (Ortiz, 2005).

Una vez a cargo de la dirección del Instituto de Física Loyarte logró atraer hacia las áreas de su interés a una parte substancial de sus miembros; podría citarse entre ellos a Margrete Heiberg de Bose, Florencio Charola, A. Fonseca, Rafael Grinfeld, Adolfo T. Williams, y otros. Loyarte quedó también a cargo de dos cátedras de Física que eran fundamentales para la formación de los jóvenes estudiantes de ingeniería y de física. La primera era el curso de Física General, obligatorio para los estudiantes de esas dos carreras. Entre fines de la década de 1920 y mediados de la siguiente, Loyarte redactó un tratado de *Física General*, diseñado especialmente para esos alumnos y fielmente adaptado a sus cursos universitarios. Su obra tuvo una influencia considerable en la enseñanza de la física básica a nivel universitario, tanto en Argentina como en otros países de la América Latina (Gangui y Ortiz, 2022.2).

La segunda de las cátedras dictadas por Loyarte, Física Teórica, era clave para la formación específica de los estudiantes de física: en este último curso Loyarte enseñó mecánica cuántica y relatividad a una generación de físicos argentinos. Uno de sus exalumnos, el profesor Mario Bunge (1919-2019), que asistió tanto a sus clases de Física General como a las de Física Teórica, ha indicado recientemente a los autores[2] que, a su juicio, el curso de Física General era excepcional. Asimismo, ha indicado que, en sus tiempos de estudiante, en el curso de Física Teórica Loyarte se ocupó de la teoría de la relatividad; a su juicio en la Argentina de esos años Loyarte era uno de los físicos que habían estudiado esa teoría con mayor profundidad.

Además de su enseñanza de la disciplina, las contribuciones científicas originales de Loyarte cubrieron temas relacionados con el estudio del magnetismo y luego de la nueva física cuántica, perspectivas que, como hemos dicho, Gans había introducido en la Argentina (von Reichenbach, 2009). Finalmente, como resultado de sus investigaciones experimentales, Loyarte propuso también una innovación audaz en el campo de la mecánica cuántica, cuya recepción fue controvertida. En las últimas secciones de este trabajo y, con mayor detalle, en una continuación de nuestro estudio (Gangui y Ortiz, 2022.1), nos ocuparemos de esos trabajos y de su impacto.

## 2. Esfuerzos para la promoción de las ciencias en la Argentina en la década de 1920: el rol de la Academia de Ciencias de Buenos Aires

### 2.1 Holmberg y la Academia de Ciencias de Buenos Aires

Loyarte fue incorporado a la Academia Nacional de Ciencias Exactas, Físicas y Naturales de Buenos Aires en el mismo año en que ascendió a la dirección del Instituto de Física. El acto de incorporación efectiva tuvo lugar el 22 de junio de 1925 en una sesión especial que contó con la asistencia del presidente de la República, Dr. Marcelo T. de Alvear (1868-1942) (Dassen, 1928, pp. 110-12). Ese fue un año memorable para la física en la Argentina: sólo un par de meses antes Albert Einstein había realizado una visita científica a ese país: desde luego, también él fue incorporado a la Academia (Ortiz, 1995).

Como discurso de incorporación a la Academia Loyarte presentó una deducción estadística de la ley de distribución de Planck, un descubrimiento íntimamente ligado a los orígenes de la mecánica cuántica (Academia, 1928.3). En ese trabajo, que había sido publicado en la revista *Contribución al estudio de las ciencias físicas y matemáticas*[3] dos años antes (Loyarte, 1926.1), Loyarte declaró que la suya era una deducción 'mucho más sencilla que la de Debye y que la de Bose'[4] (ver también Academia, 1928.4).

---

[2] Entrevista realizada en Buenos Aires en septiembre 2015.

La presencia de Alvear en ese acto público fue el resultado de los esfuerzos realizados por el presidente de la Academia, Dr. Eduardo L. Holmberg (1852-1937) a lo largo de la década de 1920 para promover esa nueva institución. Este destacado naturalista intentaba persuadir a las más altas autoridades nacionales de la importancia cultural, pero también económica, que podía redundar al país el desarrollo de la investigación científica original. En particular, era partidario de extender su profesionalización a áreas no tradicionales, como entonces era la física.

La importante acción de Holmberg en favor de la instauración y profesionalización de la investigación científica original en la Argentina en la década de 1920, lo mismo que los esfuerzos de la SCA hacia fines de la Primera Guerra Mundial, permiten comprender más cabalmente los interesantes desarrollos que tuvieron lugar en la década siguiente y que culminaron con la creación de la *Asociación Argentina para el Progreso de las Ciencias* (AAPC). Sin embargo, el grupo de investigadores que alentó esos esfuerzos en la década de 1930 es diferente al que consideramos en este trabajo. Veremos más adelante que Loyarte, también en la década de 1930, contribuyó positivamente al proceso de instauración y profesionalización de la investigación científica en el medio universitario argentino (Ortiz, 2015, pp. 39-43).

## 2.2 Premio Nacional de Ciencias y la promoción de la investigación mediante becas externas

La instauración de un Premio Nacional en el área de las Ciencias (Ley 9141, del 14 de febrero de 1925, en Academia, 1928.2) fue un nuevo elemento en la laboriosa construcción de una imagen moderna de la investigación científica en la Argentina, tarea que ocupó una buena parte de la década de 1925 a 1935 y en la que la figura singular de Holmberg también jugó un papel central.

A propuesta de la Academia, el 23 de julio de 1927 Loyarte recibió el nuevo Premio Nacional de Ciencias para 1926 (Academia, 1928.2). Ese premio era otorgado por el Poder Ejecutivo, de acuerdo con el Ministerio de Instrucción Pública y con el asesoramiento de la Academia de Ciencias. Loyarte lo compartió con el físico químico Horacio Damianovich (1883-1959), discípulo del profesor francés antes nombrado Camilo Meyer.

En esos años Loyarte era, sin duda, el único miembro de esa Academia dedicado exclusivamente a la investigación en física moderna. Una vez elegido para integrar la Comisión de Física de la Academia, fue prontamente invitado a integrarse también a la Comisión de Matemáticas y, poco más tarde, fue elegido secretario de esa misma corporación. Desde entonces su ascenso dentro de la comunidad científica de esos años fue verdaderamente meteórico.

A principios de abril de ese mismo año de 1927 Loyarte envió una carta a sus colegas del Instituto de Física invitándolos a participar en una reunión que se celebraría en una sala de ese mismo Instituto el 12 de ese mes. En esa reunión especial se discutiría la posibilidad de crear una *Junta de Ampliación de los Estudios de Ciencias Físico Matemáticas*.

Loyarte creía que con el apoyo de 'jóvenes egresados distinguidos, que gozan de posiciones oficiales [dentro y fuera de la UNLP], con el apoyo de la Universidad, con cuotas de adherentes y otros recursos que puedan lograrse, es posible mantener una organización que envíe a Europa, en forma sistemática, a los egresados que hayan evidenciado cualidades especiales de capacidad'.[5] Si bien no parece que esta interesante iniciativa haya conducido a resultados concretos señala, sin duda, el profundo impacto de las iniciativas de la antes nombrada *Junta para Ampliación de Estudios e Investigaciones Científicas* (JAE) sobre las ideas de los investigadores argentinos. La JAE había sido creada en Madrid en 1907 por iniciativa de Santiago Ramón y Cajal (1852-1934): invocando el nombre de la Junta española, Loyarte expresaba su deseo de emular a ese prestigioso consejo de investigaciones científicas en, por lo menos, una de sus actividades: el envío de graduados destacados de la UNLP al exterior.

---

[3] Esta era una revista oficial de la Universidad Nacional de La Plata. La anotaremos como *Contribución* en lo que sigue.

[4] Loyarte se refiere al trabajo del físico hindú Satyendra Nath Bose (Bose, 1924).

### 2.3 Loyarte, presidente de la Universidad Nacional de La Plata

También en 1927, a principios del mes de noviembre y sólo dos años después de su designación como director del Instituto de Física, como Académico y de recibir el Premio Nacional de Ciencias, la Asamblea Universitaria eligió a Loyarte presidente de la Universidad Nacional de La Plata. Esa designación, por un período de tres años, de 1928 a 1930, tenía una significación adicional: por primera vez en la corta historia de la UNLP se designaba a uno de sus propios exalumnos para conducirla (*El Argentino*, 1927.1).

La elección se realizó el 10 de noviembre de 1927: el presidente anterior Benito Nazar Anchorena (1884-1970), que había ocupado el cargo por dos períodos entre 1921 y 1927, intentó, sin éxito, ser reelegido; luego de algunas alternativas Loyarte obtuvo 84 votos, seguido por Tomás Casares (1895-1976) con 57, Ricardo Levene (1885-1959) con 31, Federico Walker con 27, y Nicolás Besio Moreno (1879-1952) con sólo 4.

La transmisión de la Presidencia se realizó el 1 de diciembre en el Salón de Actos del Colegio Nacional de La Plata, que dependía de la Universidad. Prudentemente, Loyarte declinó hacer declaraciones generales sobre sus planes futuros. Sin embargo, hizo afirmaciones de considerable interés: indicó que continuaría con la política educacional introducida por Joaquín V. González en la Universidad. Es decir, que trataría de dar un sentido de modernidad a la enseñanza universitaria y un papel prominente a las ciencias. Además, en un momento en el que la generación de *profesionales* parecía ser el objetivo de las instituciones de alta cultura, Loyarte insistió en la importante función que debía jugar la ciencia en la educación de profesionales: 'aun cuando [la enseñanza profesional] no va dirigida a impulsar el progreso de la ciencia', 'sino a la utilización de ésta en provecho de la sociedad, [esa enseñanza] no puede ser sino eminentemente científica' (*El Argentino*, 1927.2). En ese momento de la Argentina, y en esa universidad, sus declaraciones eran elocuentes.

Sin olvidar sus antiguos intereses por la 'ampliación de estudios', y luego de una obligada referencia a los éxitos de Ramón y Cajal y su *Junta* en España, Loyarte afirmó que 'no hay que olvidar que el problema fundamental [de la educación] consiste en asegurar la evolución de los mejores cerebros que salen de la Universidad. Su resolución la da el envío sistemático de estudiantes distinguidos a los mejores centros del extranjero. Concibo, con la evidencia fantástica del iluminado, que es posible formar así, muy rápidamente, un núcleo de sabios argentinos en las ciencias puras y aplicadas' (*El Argentino*, 1927.2).

En el terreno específico de la política nacional Loyarte era un antiguo miembro del Partido Conservador, al que había ingresado, posiblemente, en la segunda mitad de la década

---

[5] Loyarte, Director, al personal del Instituto de Física (sobre la creación de una junta de ampliación de estudios en el Instituto). Abril 7, 1927 (Ortiz, 2021).

de 1910. En el campo de la política universitaria había expresado ya, con claridad, que personalmente no estaba de acuerdo con los resultados aportados a la Universidad por la presencia de una representación estudiantil en sus organismos de gobierno. Ésta última forma de gestión había sido introducida luego de la Reforma Universitaria de 1918, que él no favorecía. Sin embargo, durante su período como presidente de la Universidad aceptó esa realidad e hizo un esfuerzo por mantener un diálogo con los representantes estudiantiles. Con ese fin mantuvo abierto el contacto personal con las autoridades de la Federación Universitaria de La Plata (FULP), que representaba mayoritariamente a los estudiantes de su universidad; en particular, con su presidente, el joven José Katz.

Desde luego, la gravitación y la capacidad operativa de ese movimiento, y de sus representantes, no era pequeña; según ha señalado Tauber (Tauber, 2015, p. 82): '[H]acia 1927, la agresiva campaña manejada por los estudiantes platenses José Katz, Bartolomé Fiorini y Ricardo Balbín culminó con la caída de Benito Nazar Anchorena' quién, como hemos indicado más atrás, aspiraba a ser reelegido por un nuevo período. Más tarde, algunos de aquellos jóvenes alcanzaron altas posiciones en diversos partidos políticos del centro o de la izquierda.[6]

El día 3 de diciembre, inmediatamente después de acceder a la Presidencia de la Universidad, Loyarte recibió la visita de los dirigentes de la FULP, encabezados por su presidente Katz. Luego de un detallado intercambio de ideas, los estudiantes expresaron su satisfacción por la salida del antiguo presidente, Dr. Nazar Anchorena, a quién ya habían definido como un 'representante de los intereses oligárquicos', y 'se manifestaron bien impresionados por los propósitos del nuevo presidente' (*El Argentino*, 1927.3). Pocos años más tarde, bajo un régimen de gobierno militar, la actitud abierta de Loyarte, cuyas credenciales conservadoras nadie podía poner en duda, fue interpretada como un signo de debilidad.

Además, durante su presidencia de la Universidad Loyarte logró iniciar un programa de envió selectivo de graduados universitarios al extranjero, donde éstos podían realizar estudios avanzados. Uno de ellos fue su antiguo colaborador, el joven físico teórico Enrique Loedel Palumbo, que fue enviado a Alemania para perfeccionarse en física (Gangui y Ortiz, 2020).

Loyarte favoreció la creación de Departamentos dentro de las Facultades, y también de nuevos Institutos de Investigación que, en líneas generales, seguían el modelo del Instituto de Física; éste, a su vez, se había inspirado en las nuevas instituciones creadas en Alemania poco antes de la Primera Guerra Mundial. La organización departamental, de la que Loyarte se sentía justamente orgulloso de haber contribuido a introducir en la Argentina, sólo fue retomada por otras universidades locales unos 25 años más tarde. Particularmente por la nueva Universidad del Sur, en Bahía Blanca, y más tarde por algunas Facultades de la Universidad de Buenos Aires.

Sin duda las responsabilidades administrativas, particularmente difíciles en esos años, tuvieron un impacto considerable sobre la vida científica de Loyarte que era, entonces, un investigador joven en plena actividad. Hemos señalado ya que al ser invitado a asumir la Presidencia de la UNLP no había cumplido aún cuarenta años.

En septiembre de ese importante año de 1927, tres meses antes de que su designación como presidente de la UNLP fuera efectiva, o que siquiera pudiera preverse, Loyarte entregó a la Secretaría de su Facultad, para ser publicado en la revista *Contribución*, el manuscrito de un estudio que se encuadraba dentro de la misma línea de investigación de otros trabajos suyos publicados anteriormente. Apareció en el tomo cuarto de esa revista, en enero de 1928 (Loyarte, 1928). Ese estudio, y otros que lo siguieron dentro de la misma línea de trabajo, le ocasionarían serios disgustos, como veremos a continuación.

---

[6] Ricardo Balbín se unió a la Unión Cívica Radical, de la que llegó a ser su máximo dirigente nacional; Bartolomé Fiorini fue un destacado profesor de Derecho; José Katz, graduado simultáneamente en Ingeniería y en Derecho, se incorporó al Partido Comunista; además, como su amigo Ernesto Sábato, fue un ajedrecista destacado.

## 3. El rol de Loyarte en el desarrollo de las investigaciones del Instituto de Física de La Plata

### 3.1 Los mecanismos de comunicación nacional e internacional empleados por los físicos de la UNLP

Localmente, Loyarte y sus colaboradores difundían los resultados de sus investigaciones en la revista universitaria platense antes nombrada: *Contribución al estudio de las ciencias físicas y matemáticas* (que hemos llamado *Contribución*). Esa publicación continuaba un esfuerzo poco conocido de Bose: la revista *Mitteilungen aus dem physikalischen Institut der National Univeristät La Plata*[7] que, históricamente, sería la primera revista dedicada exclusivamente a la física en la Argentina. *Mitteilungen* comenzó a publicarse en 1911: el No. 1 de esa serie es una nota de E. Bose y el No. 2 un trabajo conjunto de M. H. de Bose y E. Bose.[8] A causa del inesperado fallecimiento de Bose, esa serie se detuvo con ese segundo número (Ortiz, 2021).

*Contribución*, la continuación de *Mitteilungen*, fue obra de Gans: alcanzó un éxito y una circulación considerablemente más amplia que su fugaz antecesora. En 1935 era suficientemente robusta como para ser dividida en dos series, cada una de ellas apuntaba a una audiencia diferente: *Serie Física* y *Serie Matemática*. El desarrollo de esta última revista ha sido considerado en (von Reichenbach, 2007).

Algunos de los trabajos publicados en *Contribución*, particularmente aquellos que mostraban una mayor originalidad, eran también sometidos para su publicación, y a menudo aceptados, en revistas científicas de circulación internacional. Para facilitar este proceso, previamente se vertía su texto a un idioma extranjero: al alemán cuando el trabajo era enviado, por ejemplo, a *Physikalische Zeitschrift*, o al francés cuando el objetivo era publicarlo, por ejemplo, en el *Journal de Physique et le Radium*. El texto de esas versiones era generalmente idéntico, o casi idéntico, al publicado en español en *Contribución*. Puede decirse que las versiones en idioma español estaban principalmente dirigidas a informar al medio local acerca de la laboriosidad de sus autores; las versiones en alemán o francés tenían la intención de lograr una inserción más amplia en el movimiento científico contemporáneo.

*Physikalische Zeitschrift* tenía una conexión muy especial -y muy importante en relación con nuestro trabajo- con el Instituto de Física de La Plata: hasta su partida a la Argentina, en 1909, su redactor principal había sido Emil Bose. Cuando éste falleció en La Plata *Physikalische Zeitschrift* publicó un extenso artículo, redactado por su esposa Margrete (Heiberg de Bose, 1911), en el que se describían sus ricas instalaciones, obra de Bose. Más tarde, entre los colaboradores platenses de esa revista se contó Loyarte y, también, algunos de sus colaboradores: Charola, Grinfeld, Heiberg de Bose, Loedel Palumbo, Williams y otros.

---

[7] *Comunicaciones del Instituto de Física de la Universidad Nacional de La Plata*.
[8] En carta fechada el 23 de abril de 1940 y dirigida al Ing. Juan Sábato, que estaba redactando una nota sobre las publicaciones de la UNLP, Margrete Bose le señaló la existencia, en 1911, de *Mitteilungen aus dem physikalischen Institut der National Univeristät La Plata* (Ortiz, 2021).

## 3.2 Física experimental y mecánica cuántica

En las primeras décadas del siglo XX diversos grupos de investigadores en Europa y en forma creciente también en los Estados Unidos, intentaron verificar experimentalmente la validez de las ideas que se derivan del modelo de la estructura del átomo propuesto por Bohr. La existencia de niveles discretos de excitación fue demostrada, experimentalmente, por James Franck (1882-1964) y Gustav Hertz (1887-1975) en 1913 empleando un dispositivo experimental similar al que, con anterioridad, había desarrollado Philipp Lenard (1862-1947) (White, 1934). Esquemáticamente, ese equipo consistía en un tubo cerrado, que contenía cuatro electrodos y al que se le inyectaba vapor de mercurio,[9] o de otros elementos; el mercurio en fase gaseosa se prefería por su electronegatividad: no rechazaba naturalmente a los electrones incidentes. Los electrodos eran: un cátodo, que calentado por un filamento podía emitir electrones (sistema de excitación); una o dos grillas cargadas positivamente, capaces de acelerar a esos electrones en su tránsito hacia una segunda placa (sistema de detección); finalmente, existía la opción de interponer filtros (por ejemplo, de cuarzo, calcita u otros materiales.[10] Por esos trabajos Franck y Hertz recibieron el Premio Nobel de Física en 1925. En el tomo IV de su *Física General* Loyarte describió los dispositivos de Lenard-Frank-Hertz (Loyarte, 1935, p. 508). En alguno de sus trabajos de esos mismos años, por ejemplo, en (Loyarte y Heiberg de Bose, 1935, p. 26), Loyarte hizo referencia a la descripción que había dado en su *Física General*.

Con esos equipos se trataba de verificar si las transferencias de energía causadas por colisiones inelásticas de electrones libres con átomos de un elemento específico[11] (el vapor que llenaba el dispositivo) ocurrían o no en forma discreta, es decir, si se presentaban revelando diferentes niveles, en forma *cuantificada*, tal como postulaba la teoría de Bohr.

Franck y Hertz encontraron que la corriente de los electrones que se movían desde la placa caliente de uno de los extremos (el cátodo) hacia el otro (el ánodo) sufría una caída cuando el campo acelerador alcanzaba 4,9 volts, y que lo mismo se repetía con múltiplos de esa cantidad. Esto indicaría que, con aquellas intensidades de campo eléctrico (que era aquello que impartía la aceleración de los electrones), átomos de gas contenidos dentro de la ampolla efectivamente habían absorbido energía cinética de los electrones. En este experimento la variable independiente era la energía cinética de los electrones (la tensión aplicada para acelerarlos) y, la dependiente, los valores de la interacción energética (la corriente detectada en la segunda placa).

En la transición de la década de 1910 a la de 1920 varios investigadores, incluyendo a Franck, continuaron estudiando estos problemas y trataron de determinar, metódicamente, los estados de energía (o *potenciales críticos*) a los que se podía elevar el gas de mercurio mediante el impacto con electrones. En (Franck y Einsporn, 1920) Franck, continuando esos estudios con la colaboración de su entonces alumno Enrich Einsporn (1890-1964), trató de medir la energía residual de los electrones después de acelerarlos con un campo eléctrico en una región que contenía átomos de un determinado gas o vapor: corrientemente vapor de mercurio por las razones dadas más atrás. Los intercambios críticos de energía se podían medir de diferentes maneras: por la energía residual de los electrones o, también, por la radiación emitida por átomos a los que las colisiones inelásticas los hubieran excitado a un nivel superior; esto último se podía detectar con medios espectroscópicos o fotoeléctricos.

---

[9] Este aparato, originalmente debido a Franck y Hertz, permitía obtener una mayor resolución y consecuentemente detectaba un número mayor de picos de tensión.
[10] Como efectivamente lo hizo Helen Messenger, en (Messenger, 1926).
[11] Para analizar esos intercambios de energía se estudiaban, por ejemplo, posibles cambios en la distribución de velocidades de un paquete de electrones.

El objetivo final de esas investigaciones era, obviamente, resolver un difícil problema inverso: inferir aspectos de la estructura, o condiciones de estabilidad, del átomo de un elemento dado mediante una herramienta experimental basada en el estudio de colisiones (o de la lectura de líneas espectrales), comparando luego esos resultados con las previsiones de la teoría.

**3.3 Las investigaciones de Loyarte y sus colaboradores en la segunda mitad de la década de 1920**

Como hemos señalado, en la década de 1920 tanto las mediciones eléctricas delicadas como la espectroscopía eran técnicas experimentales en las que los antiguos alumnos de Gans habían logrado reunir cierta experiencia. Ambas jugarían un papel central, aunque no exclusivo, en los trabajos de investigación de Loyarte. En esos años el análisis de la compleja multitud de líneas en el espectro de una muestra parecía ofrecer un posible camino para comprender más cabalmente la estructura del átomo emisor; Loyarte siguió ese camino. Entre sus colaboradores jóvenes en el campo de la espectroscopía Adolfo T. Williams era una figura sobresaliente, de quién Loyarte recibió considerable apoyo.[12]

Interesa agregar también que, en esos años, además de su obvio interés científico por esclarecer la estructura fundamental de la materia, en los países técnicamente más avanzados había objetivos tecnológicos muy concretos en el estudio de las descargas en gases. Por ejemplo, lograr un diseño más eficiente de las válvulas termoiónicas utilizadas entonces en los equipos de radio comunicaciones y, también, en otras aplicaciones técnicas (Meek y Craggs, 1953). Esas motivaciones faltaban en la Argentina.

El propio equipo de Franck y Einsporn que, como hemos dicho, consistía esencialmente de un ánodo caliente, un cátodo y una o más grillas interpuestas entre ellos, puede pensarse como una realización experimental de un tipo particular de las nuevas válvulas electrónicas (White, 1934). Desde luego, Loyarte no era ajeno a esa realidad: en alguno de sus experimentos empleó, él mismo, componentes de equipos de radio telefonía (ver, por ejemplo, Loyarte, 1929.1, pp. 501-502).

Loyarte muy probablemente participaba del grupo internacional de *investigadores experimentales cuánticos* que, hacia fines de la década de 1910 o principios de la de 1920, trataban de encontrar una explicación para una serie de valores anormales observados en los potenciales de excitación del átomo de mercurio. Como en otros laboratorios del mundo,[13] Loyarte y su grupo utilizaron una variedad de técnicas experimentales; sus referencias teóricas eran los resultados contemporáneos de la mecánica cuántica.

Loyarte mismo contribuyó a difundir las nuevas ideas de la mecánica cuántica en la Argentina: primero a un nivel general en su trabajo de *Contribución* (Loyarte, 1927.1) y luego incorporando esa disciplina en la enseñanza universitaria, a través de sus cursos avanzados de Física Teórica. De este último curso de Loyarte quedan notas muy precisas tomadas por dos de sus asistentes jóvenes: el físico Rafael Grinfeld y el matemático Alberto Sagastume. Más tarde, con el apoyo generoso de Loyarte, que en esos años era editor de los *Anales de la SCA* (ASCA) para el área de la Física, Grinfeld y Sagastume publicaron sus notas en ASCA, bajo su propio nombre, (Sagastume Berra y Grinfeld, 1928-1930); luego en forma de libro en (Sagastume Berra y Grinfeld, 1930).

---

[12] Cuando Loyarte ocupó cargos políticos fuera de la Universidad, encargó a Williams secundarlo como director interino del Instituto de Física.

[13] En (Loyarte y Heiberg de Bose, 1935) se da una lista amplia de nombres de científicos trabajando en esa línea de investigación, entre ellos: E. O. Lawrence y T. C. Morris, aunque sin citar específicamente sus trabajos que, posiblemente, en esos años eran familiares para sus lectores.

En 1926, siguiendo la forma de doble comunicación a la que hemos hecho referencia más atrás, Loyarte publicó en *Contribución* una nota en la que informaba sobre los resultados de sus estudios acerca de los potenciales de excitación del átomo de mercurio (Loyarte, 1926.2).[14] Prácticamente el mismo texto apareció en alemán en *Physikalische Zeitschrift* (Loyarte, 1926.3). Ese doble trabajo consolidaba una línea de investigación que Loyarte había iniciado a comienzos de la década de 1920, es decir, aun antes de la partida de Gans, en la que como hemos señalado ya se trataba de estudiar, en considerable detalle, posibles regularidades en datos obtenidos experimentalmente para el mercurio. Loyarte encontró que algunos de esos datos, obtenidos para el mercurio y para otros elementos químicos, por él o por otros investigadores, no encajaban claramente con las predicciones de la teoría. A continuación veremos que ese germen de idea lo condujo a proponer una explicación radical, que implicaba atribuir una significación especial al potencial de 1,4 volt y, posteriormente, a la introducción de un nuevo número cuántico.

### 3.4 Loyarte, la Academia de Ciencias de Buenos Aires y el potencial de adición de 1,4 volts

También en 1926, la Academia de Ciencias de Buenos Aires resolvió invitar a sus miembros a presentar comunicaciones sobre las investigaciones científicas originales que estuvieran realizando. Loyarte fue uno de los primeros en ofrecer un resumen de los resultados de sus estudios recientes (ver Dassen, 1928, pp. 119-120). En la sesión de la Academia del 19 de junio de ese año Loyarte leyó una comunicación en la que se ocupó de un número de potenciales de excitación del átomo de mercurio para el cual no existía una explicación satisfactoria (Academia, 1928.3). Señaló que 'ellos ponen de manifiesto que los cinco potenciales de excitación, a los cuales no correspondían líneas ópticas ni términos de serie, provienen de la existencia de un único potencial de adición. Con esto se explica de un modo muy satisfactorio las medidas de Franck y Einsporn y las del conferenciante. Con esta base, y en colaboración con el doctor Adolfo Williams, se procedió a inquirir si no existirían series anormales. Los resultados de este estudio hacen muy probable la existencia de las mismas, que provendrían de un átomo inestable'.

*Anales* publicó una versión francesa de ese trabajo de 1926: (Loyarte, 1929.1), que el matemático Claro Dassen (1873-1941) tradujo especialmente para Loyarte. Ese mismo artículo ya había aparecido: en español en *Contribución* (Loyarte, 1926.2) y, en alemán, en *Physikalische Zeitschrift* (Loyarte, 1926.3), ambos en 1926.

Loyarte hizo referencia a un *potencial de adición de 1,4V*, del que veremos más detalles en lo que sigue (Dassen, 1928, p. 113). Interesa señalar que, en ese momento particular en la vida de la joven Academia de Ciencias de Buenos Aires no era frecuente escuchar de sus miembros contribuciones que reflejaran un nivel de conversación tan fluido con la comunidad científica internacional como sugerían las publicaciones de Loyarte.

También dentro de 1926, en la reunión de la Academia del 18 de noviembre, Loyarte presentó otro trabajo: 'Las presuntas series anormales en los potenciales de excitación del átomo de mercurio' (Academia, 1928.1), que había sido hecho en colaboración con Williams, cuyo texto final apareció muy poco tiempo después, en 1927 y con un título más breve, en *Contribución* (Loyarte y Williams, 1927).

A fines del año siguiente, el 17 de septiembre de 1927, Loyarte volvió a informar a la Academia sobre sus investigaciones exponiendo trabajos realizados por él mismo o en colaboración con Williams. En el trabajo con este último discutieron, nuevamente, la 'Rotación cuántica del átomo de mercurio' (Loyarte y Williams, 1928), ampliando así trabajos antes citados, que Loyarte había publicado en 1927 en *Physikalische Zeitschrift* (Loyarte, 1927.2) o que aparecerían en fecha muy próxima, en enero de 1928, en *Contribución* (Loyarte, 1927.3); ver también la referencia a Loyarte en (Dassen, 1928, pp. 119-120).

---

[14] En conexión con este trabajo, ver también (Williams y Loyarte, 1926).

En su comunicación a la Academia Loyarte indicó que 'la sospecha de la existencia de tal proceso fue sugerida por la aparición de potenciales de adición [que son] múltiplos de un mismo potencial (1,4 voltios)', según resultaba de sus medidas, de las de Franck y de las de Jarvis.[15] Las investigaciones ópticas por él efectuadas '*comprueban plenamente la existencia de una rotación cuantificada en el átomo de mercurio, hecho enteramente desconocido* [las itálicas son nuestras]. Además, quedó evidenciado otro hecho nuevo: que los cuántas de rotación, tanto se suman a las energías de los saltos cuánticos de los electrones como se restan de ellos. La revelación de estos hechos, y de una nueva forma de combinación de las energías cuantificadas de los electrones y de la rotación, facilitará el estudio de los espectros y, por consiguiente, de la mecánica de los átomos y moléculas' (ver Dassen, 1928, pp. 119-120). Era un anuncio extraordinario, que difícilmente los miembros de esa corporación, en su gran mayoría ajenos a la Física, podían quizás abarcar en toda su magnitud.

**3.5 El impacto internacional de los trabajos de Loyarte**

En esos años el físico V. Pavlov dirigía un grupo de investigadores en el Laboratorio Nacional de Física y Tecnología de Leningrado, en la URSS; ese investigador era entonces uno de los más activos en el área en la que también trabajaba Loyarte. Junto con su alumna N. Sueva, Pavlov señaló en 1929 (Pavlov y Sueva, 1929) que si se representa el número relativo de electrones lentos en función de una tensión V que los acelera se encuentra una curva que presenta un número de máximos, cuya posición es útil para determinar las tensiones críticas. Indicaron también que, por encima de 4,66 volts, los valores de esa serie mostraban resultados ya conocidos acerca de la tensión crítica del vapor de mercurio; para ellos se conocían las líneas espectrales correspondientes. Sin embargo, esos autores señalaron que aquellos valores incluían algunos voltajes, por ejemplo 5,25, 5,75, 6,05, 6,30, 7,10, 7,45 y 8,05 volts, para los que no se conocían las líneas espectrales correspondientes. Indicaron también que un número significativo de tensiones críticas se encontraban dentro del rango de 0,45 a 0,466 volt, que está debajo del primer nivel de estimulación del átomo normal de mercurio.

No dejaron de señalar que existían dificultades para interpretar los valores experimentales e hicieron una referencia explícita a los resultados aportados por Loyarte que, como hemos indicado, habían sido difundidos internacionalmente por *Physikalische Zeitschrift*. Pavlov y Sueva señalaron que: 'Loyarte también ha determinado, experimentalmente, tensiones críticas para el vapor de mercurio' (en Loyarte, 1926.3). Con referencia a sus propuestas teóricas agregaron que el investigador platense 'es de la idea de que constituye una energía de rotación de los átomos' (Pavlov y Sueva, 1929, p. 245).

---

[15] Se refiere a Charles W. Jarvis, un pionero en los estudios de la radiación correspondiente a potenciales críticos en el vapor de mercurio (Jarvis, 1926).

## 3.6 La segunda mitad de década del 1920: las visitas de Einstein y Langevin a la Argentina

Más atrás nos hemos referido a los esfuerzos de la *Institución Cultural Española* desde mediados de los años 1910, que llevó a la Argentina a un grupo sobresaliente de intelectuales y científicos de aquel país (Ortiz, 1988). Esos esfuerzos fueron retomados en 1922 por *Die Deutsche-Argentinische Kulturgesellschaft* y culminaron con la visita de Albert Einstein. Más tarde, el *Institute de l'Université de Paris à Buenos Aires* se unió a ese grupo de asociaciones propiciando, entre otras, la visita del profesor Paul Langevin (1872-1946) a la Facultad de Ciencias de Buenos Aires. En esos años Langevin era uno de los físicos más destacados de su generación: miembro del *Collège de France*, de la *Académie des Sciences*, profesor en la *École Supérieur de Physique et Chemie* y redactor principal de una prestigiosa revista francesa de física: el *Journal de Physique et le Radium.*

Durante su visita de 1928 Langevin dictó, en las mismas aulas de la Facultad de Ciencias de la Universidad de Buenos Aires donde tres años antes Einstein había desarrollado su famoso ciclo de conferencias (Ortiz, 1995; Gangui y Ortiz, 2008), un curso sobre la Teoría de la Relatividad que estaba específicamente dirigido a los estudiantes de ciencias Físico Matemáticas de esa universidad. La prensa local reseñó puntualmente el contenido de cada una de sus exposiciones, como lo había hecho antes con las de Einstein (Academia, 1929) y con las de otros visitantes ilustres.

A pedido de Loyarte, Langevin visitó también el Instituto de Física de la Plata; en esa ocasión su director tuvo oportunidad de mostrarle las espléndidas instalaciones de ese Instituto y referirse a los temas que, en ese momento, atraían la atención de varios de sus miembros. Desde luego, se refirió a las ideas que él mismo había desarrollado en trabajos publicados recientemente en *Physikalische Zeitschrift*, que Langevin difícilmente podía ignorar.

Muy poco después de ese encuentro, en un trabajo ya citado (Loyarte, 1928), publicado en el mismo año de la visita de Langevin a la Argentina, Loyarte hizo referencia a que había comunicado los resultados de su memoria a ese distinguido físico 'en ocasión de su vista al Instituto de Física y por carta de fecha septiembre 4'. Indicaba también que, muy gentilmente, Langevin había hecho, específicamente para él, uno de los cálculos consignados en su memoria (Loyarte, 1928, p. 228). Loyarte no precisó si Langevin había respondido a su carta, o había hecho algún comentario concreto sobre sus resultados, o sobre la dirección general de sus estudios.

Cabe señalar, también, que en el año de la visita de Langevin a Buenos Aires Loyarte y Williams publicaron, en *Contribución* y también en *Physikalishes Zeitscrhrift,* resultados de sus investigaciones sobre espectroscopía experimental que han resistido el paso del tiempo y continuaron siendo citados; por ejemplo (Loyarte y Williams, 1929.1).[16]

Interesa señalar que las ideas políticas de Loyarte diferían sustancialmente de las de Langevin: el primero era un convencido conservador, con una visión paternalista de su rol en la sociedad, mientras que el segundo era un conocido pacifista que, en Francia, se había acercado al socialismo. A pesar de esas obvias diferencias, fue Loyarte quien impulsó la incorporación de Langevin como Miembro Honorario de la Academia Nacional de Ciencias Exactas, Físicas y Naturales de Buenos Aires, como una celebración de su prolífica visita a la Argentina. Loyarte presentó su propuesta el 28 de Julio de 1928, la Academia la aceptó el 11 de agosto y el 25 del mismo mes tuvo lugar la ceremonia de incorporación de Langevin (Academia, 1931); el interesante discurso de Loyarte se reproduce en páginas 332-336 de esa publicación.

---

[16] Citado, por ejemplo, en tablas de resultados sobre los niveles de energía en el átomo, derivados del análisis de espectros ópticos. Ver, por ejemplo, (Moore, 1971, II, p. 97).

Como requerían los reglamentos de la Academia, la iniciativa de Loyarte fue avalada con la firma de doce académicos, todos ellos bien conocidos en el campo específico de sus estudios y, también, en los círculos científicos del país. Sin embargo, sólo Loyarte y el físico químico Damianovich,[17] a quién nos hemos referido más atrás, eran los dos únicos científicos cuyos estudios les permitían apreciar la importancia de la obra científica de Langevin. En esos años ninguno de los físicos de la Facultad de Ciencias de Buenos Aires era aún miembro de esa Academia.

Precisamente en los años en que visitó la Argentina, Langevin había comenzado a explorar, también, el rol de la ciencia pura en el adelanto de la sociedad humana. Luego desarrollaría ese tema, con notable originalidad, en diversas publicaciones francesas. Durante su visita a Buenos Aires dictó una conferencia pública sobre ese tema, titulada 'Las funciones sociales de la investigación científica', que tuvo un impacto amplio y considerable. El diario *La Prensa* la hizo traducir y la incluyó, *in extenso*, en sus páginas (Langevin, 1928).

## 4. Recepción de la rotación cuantificada fuera de la Argentina

### 4.1 Las revistas internacionales de referencia

A continuación, reseñaremos muy brevemente, la recepción que tuvieron, en el mundo científico nacional e internacional de esos años, los trabajos de Loyarte y de sus colaboradores platenses sobre la rotación cuántica del átomo de mercurio. Para estudiar ese proceso haremos uso de una guía que es, a la vez, novedosa y contemporánea: las notas publicadas en algunas de las revistas internacionales de reseña científica más altamente calificadas de esos años.

Evaluar por este medio la crítica local es más difícil, ya que en la Argentina de la primera mitad del siglo XX ese género de revistas verdaderamente no existía para las ciencias; menos aún para la física. Como es sabido, las secciones de bibliografía de ASCA, o de otras revistas científicas argentinas de similar envergadura, daban preferencia a la recensión de *libros* más bien que a la de artículos científicos originales.[18] Sin embargo, como veremos en (Gangui y Ortiz, 2022.1), hubo una excepción local que, afortunadamente, incluye a algunos de los artículos de Loyarte a los que nos referimos en este trabajo.

Para el extranjero utilizaremos tres de las revistas de reseña más valoradas en esos años que, además, corrientemente hacían referencia a trabajos de Loyarte y sus colaboradores. La primera de ellas es *Chemical Abstracts*, la segunda es *Science Abstracts - Section A - Physics Abstracts*[19] (más conocida como *Science Abstracts – Physics*, o *Physics Abstracts*), y la tercera *British Chemical Abstracts*; las tres tenían entonces una audiencia internacional muy amplia. La primera se publicaba en los Estados Unidos, las dos últimas en Inglaterra. A pesar de que dos de esos periódicos eran, formalmente, revistas de referencia sobre estudios en el campo de la química, ambas contenían una sección, frecuentemente designada *Pure Chemistry*, en la que se reseñaban trabajos de físico química (también llamados de *Química teórica*) y de física moderna. En esos años la física carecía del fuerte apoyo económico que la química podía reclamar de la industria. Puede decirse que las dos revistas de reseña de artículos de química a las que haremos referencia cubrían, tradicionalmente y con suficiente amplitud, las áreas en las que trabajaban Loyarte y sus colaboradores platenses.

---

[17] Que había ingresado a esa corporación en 1916.
[18] Los comentarios de libros son, sin embargo, un documento de considerable importancia que nos permite adquirir una idea bastante precisa de lo que se leía y de lo que se valoraba en diferentes períodos históricos del Buenos Aires científico.
[19] Formalmente, desde 1898, sus editores eran *The Physical Society* y *The Institution of Electrical Engineers* (IEE). Ambas instituciones estaban radicadas en Londres.

Finalmente, es importante recordar que las reseñas que aparecían en diferentes *Review-journals* publicados en el extranjero sólo intentaban *informar* a sus lectores, tan objetivamente como fuera posible, sobre el contenido del trabajo reseñado. Generalmente no hacían una evaluación crítica; es decir, una aceptación o rechazo de los resultados reseñados. Sin embargo, en casos muy especiales, encontramos en esas reseñas opiniones contundentes: en (Gangui y Ortiz, 2022.1) citaremos algún ejemplo en conexión con los trabajos que nos interesan.

Tanto *British Chemical Abstracts* como *Physics Abstracts* registran una cobertura relativamente extensa de los trabajos científicos de físicos platenses, mayor que otras revistas de referencia; particularmente de los trabajos de Loyarte y de sus colaboradores. En parte esta circunstancia ha sido determinante en nuestra selección; importa destacar, también, que, en esos mismos años *Physics Abstracts* había decidido incluir a la revista platense *Contribución* en el fondo de revistas que reseñaba regularmente: esta circunstancia nos obliga a prestarle una especial atención. Es posible que esa decisión haya sido motivada por el hecho de que Gans, ocasionalmente, adelantaba los resultados de sus importantes investigaciones en *Contribución*, antes de que estos aparecieran en revistas europeas de física de gran circulación. Finalmente, en las páginas que siguen dejaremos de lado referencias a trabajos de Loyarte, o de Loyarte con sus colaboradores platenses, incluso Loedel Palumbo, que no tienen una relación directa con el problema de la rotación cuantificada de los átomos.

## 4.2 La recepción de la rotación cuantificada de Loyarte en revistas internacionales de referencia

En 1926, inmediatamente después de la partida de Gans a Alemania, y siguiendo la costumbre antes citada, Loyarte publicó dos trabajos muy similares: uno en español, que apareció en *Contribución* (Loyarte, 1926.2) y el otro, de hecho, su traducción al alemán, que difundió *Physikalische Zeitschrift* (Loyarte, 1926.3). Como ya señalamos, estos trabajos consolidaban una línea de investigación que el investigador platense había transitado desde el comienzo de los años 1920. Recordemos que la medición de parámetros eléctricos o espectroscópicos delicados era una de las técnicas experimentales en las que se había logrado reunir cierta experiencia en La Plata en la época de Gans: esas técnicas juegan un papel central, aunque no exclusivo, en la serie de trabajos que vamos a considerar.

En el mismo año en el que apareció publicado el trabajo (Loyarte, 1926.3), A. E. Garret lo reseñó en *Science Abstracts - Physics,* (Garret, 1926) señalando que el autor afirmaba que el mercurio podía jugar un rol anormal, lo que ya se habría observado en el caso del talio, su vecino en la tabla periódica.

El año siguiente Loyarte publicó un nuevo trabajo: (Loyarte 1927.2), que B. J. C van der Hooven (van der Hooven, 1927) reseñó para *Chemical Abstracts* indicando que su autor ofrecía evidencia experimental de que un cierto número de potenciales de excitación del mercurio (cinco de los cuales habían sido observados por Franck y Einsporn), que no correspondían a 'líneas espectrales o a diferencias conocidas',[20] se podrían deducir muy simplemente por la adición o substracción de 1,4 volts (o de un múltiplo de esa cantidad) a potenciales conocidos. El mismo trabajo, (Loyarte 1927.2), fue reseñado para *Science*

---
[20] Dice, específicamente: 'spectral lines or term differences'.

*Abstracts - Physics* por W. Stiles[21] (Stiles, 1928), que fue más lejos que su colega. Primeramente, señaló (lo mismo que van der Hooven) que luego de revisar la evidencia experimental, Loyarte afirmaba que, mediante la adición o substracción de 1,4 volts, o de un múltiplo simple de esa cantidad era posible deducir, de los potenciales correspondientes a líneas ópticas conocidas (de las series de arco), una diversidad de potenciales de excitación del mercurio. Agregó que para explicar ese número -los 1,4 volts- Loyarte proponía la existencia de una *rotación cuantificada del átomo* en la que aquella cantidad representaría, precisamente, el *quantum* que corresponde a una rotación particular. Decía, además, que Loyarte daba como ejemplo un número de líneas ópticas que confirmaban sus hipótesis. Además, Loyarte había estimado el momento de inercia del átomo de mercurio, encontrando para su radio r = $6.10^{-11}$ cm, cantidad que había comparado con la que se deduce de la teoría cinética (r = $10^{-8}$ cm) y con el resultado obtenido por Rutherford para el torio (r = $7.10^{-12}$ cm).

Finalmente, en un trabajo publicado en 1928, Loyarte y Williams (Loyarte y Williams, 1928) declararon que habiendo ya establecido Loyarte, en (Loyarte, 1926.2) y (Loyarte, 1926.3), la existencia 'de series de potenciales que difieren en 1,4 volts o bien en uno de sus múltiplos…. [y] aplicando el concepto de constantes de desplazamiento' esos dos autores habían 'clasificado esos potenciales en series espectrales'. En este trabajo estudiaron, además, el espectro del talio.

## 4.3 El descubrimiento de Loyarte y sus colaboradores: un nuevo *cuanto de energía rotacional*

En los años inmediatamente siguientes Loyarte publicó, como único autor, o juntamente con algunos de sus colaboradores, nuevos estudios en esa misma línea de investigación que corroboraban sus descubrimientos anteriores. En 1928-29 Loyarte publicó dos trabajos casi idénticos sobre la rotación cuantificada del átomo de mercurio: uno en alemán, en *Physikalische Zeitschrift* (Loyarte, 1929.4) y el otro en español, en *Contribución* (Loyarte, 1928). Stiles reseñó las dos versiones en *Science Abstracts - Physics* (Stiles, 1930.1). Indicó que en ese par de trabajos el autor insistía en afirmar que los potenciales críticos (o estados de energía) del mercurio, que no corresponden a saltos cuánticos entre niveles ópticos de la serie del arco, se pueden deducir sumando o restando un múltiplo de 1,4 volts a los que cumplen con aquella condición. Agregó que Loyarte ofrecía, nuevamente, una interpretación teórica de la naturaleza física de esta cantidad: sugería que se trataría de *un nuevo cuanto de energía rotacional* del átomo de mercurio. Stiles indicó también que, según Loyarte, esa rotación no influía sobre el valor teórico del calor específico. Finalmente, agregó que el autor había ofrecido una explicación del posible origen de la rotación cuantificada y del mecanismo de intercambio de energías entre un átomo rotatorio y un electrón incidente.

En otro par de trabajos, publicados también en 1929, en español en (Loyarte, 1929.2) y en alemán en (Loyarte, 1929.3), Loyarte extendió al talio las hipótesis que antes había propuesto para el mercurio. Stiles reseñó esos trabajos en (Stiles, 1930.2), señalando que, si se supone que se trata del valor discreto de la energía de un rotador con eje libre, definido según la mecánica ondulatoria, Loyarte había deducido el momento de inercia del átomo de talio con respecto a un eje que pasa por su centro de gravedad. Agregó que su resultado coincidía con el momento de inercia del electrón respecto de un eje que pasa por el núcleo, tal como se deduce de la teoría electrónica del magnetismo. Indicó también que, para la susceptibilidad magnética, Loyarte había utilizado el valor encontrado experimentalmente para el talio sólido (similarmente a lo que Loyarte había indicado en Loyarte, 1928, p. 228).[22]

---

[21] Stiles fue un destacado especialista en espectroscopía, pionero en el estudio del mecanismo de la visión humana, particularmente de la percepción de los colores. Durante la Segunda Guerra Mundial estudió el impacto relativo de la iluminación urbana sobre la visibilidad en relación con la posibilidad de reducir el impacto de los bombardeos, (War Cabinet Diaries, 1940). Trabajó en el *National Physical Laboratory*, Teddington, y fue elegido miembro de la *Royal Society*, Londres; además, en 1967, esa sociedad le otorgó la prestigiosa Medalla Newton.

También en 1930 Richard Alan Morton,[23] que ya había reseñado dos trabajos más convencionales de Loyarte y Williams (Loyarte y Williams, 1929.1; 1929.2), se ocupó de (Loyarte, 1929.4) en *British Chemical Abstracts* (Morton, 1930.1). Morton indicó que, según Loyarte, los potenciales registrados por varios autores (y que no corresponden a saltos cuánticos entre niveles ópticos en la serie del arco del mercurio) podrían obtenerse mediante la adición o substracción de 1,4 volts, o de uno de sus múltiplos simples, a los números de onda de la serie de líneas. Agregó que Loyarte atribuía ese resultado a la existencia de 'una rotación cuantificada del átomo de mercurio correspondiente, en energía discreta, a un rotador con eje libre según la mecánica ondulatoria'. Dijo también que, siguiendo a Loyarte, esa rotación no influía sobre el valor teórico, (5/3), del calor específico del átomo de mercurio. Morton agregó que, en ese trabajo, el autor consideraba el mecanismo de intercambio de energías durante la colisión de un electrón con un átomo (concebido éste como un rotador) y también el origen de la rotación cuantificada, cuya existencia él sostenía en el caso del mercurio.

Entretanto, en 1929, Loyarte reunió sus estudios sobre este tema en una obra de algo más de 100 páginas que editó su Universidad: (Loyarte, 1929.5).[24]

En una continuación de nuestro estudio (Gangui y Ortiz, 2022.1) veremos cómo rápidamente comienzan a surgir en el extranjero indicadores que ponen en duda los resultados experimentales y, en general, las ideas recién presentadas de Loyarte. Presentaremos también, en detalle, la polémica que surgió en el seno de su propio Instituto de Física cuando uno de los jóvenes integrantes de ese centro publica, tanto en la revista platense como en el extranjero, trabajos que muestran las debilidades de los resultados de Loyarte. Por último, describiremos la intensa actividad pública y política de Loyarte en la década de 1930 y hasta el final de su vida.

**Consideraciones finales**

La trayectoria del Dr. Ramón G. Loyarte, científico, educador y político universitario, ejemplifica a un científico argentino destacado que, muy joven, alcanzó la dirección de un instituto de investigación, el Instituto de Física de La Plata, en el que el Estado nacional había hecho una inversión substancial en recursos humanos, técnicos y económicos a lo largo de varias décadas.

En paralelo con sus tareas de investigación en física, Loyarte ocupó importantes posiciones universitarias y luego políticas. Fue, también, el primer egresado de la

---

[22] La anomalía entre los valores calculados por Langevin y los utilizados por Loyarte en (Loyarte, 1928), señalada por (Stiles, 1930.2), ha sido también observada en (von Reichenbach y Andrini, 2015).

[23] Morton, profesor en la Universidad de Liverpool, fue un pionero de la aplicación de la espectroscopía a las investigaciones bioquímicas.

[24] La Biblioteca del Maestro, en Buenos Aires, tiene un ejemplar de esta obra, dedicado por Loyarte al escritor y poeta Leopoldo Lugones, director de esa Biblioteca, que tenía un interés particular por los adelantos de las ciencias.

Universidad Nacional de La Plata que llegó a ser elegido para la presidencia de la universidad en la que se había educado. Desde esa posición auspició el desarrollo de institutos de investigación y la creación de becas externas para el perfeccionamiento de graduados en áreas nuevas de las ciencias. Aunque opuesto a la Reforma Universitaria, en el Consejo Universitario logró mantener el diálogo con figuras importantes de la oposición reformista.

En el escenario de la ciencia argentina Loyarte alcanzó también posiciones destacadas: fue elegido miembro de la Academia de Ciencias de Buenos Aires, la que más tarde le otorgó el Premio Nacional de Ciencias por sus investigaciones en física. En los primeros años de la vida de esa Academia Loyarte contribuyó con iniciativas positivas que favorecieron su desarrollo y atribuyeron a la investigación científica novedosa y a los avances más recientes de la ciencia una posición dominante dentro de sus actividades.

En el campo de la investigación científica original se destaca también su esfuerzo por impulsar una serie de trabajos emprendidos personalmente o junto con algunos de sus alumnos y colaboradores. Los resultados de esas investigaciones fueron publicados localmente y también en revistas extranjeras de gran circulación en esos años.

Sin embargo, un tópico central en las investigaciones experimentales de Loyarte, que involucraba la posible definición de un nuevo número cuántico, fue recibido con cierto escepticismo por los especialistas del extranjero. Para tener una perspectiva más amplia de la recepción de sus trabajos por parte de la comunidad internacional, nos hemos basado en los comentarios que muchos expertos publicaron en revistas específicas de reseña, cuyo objetivo era ofrecer una guía a los investigadores sobre la actualidad de diversas líneas de investigación, en una época en la que ya el número de publicaciones técnicas crecía y se volvía inabordable.